\documentclass[useAMS,usenatbib]{aa}
\usepackage{amssymb, hyperref, rotating, multirow, threeparttable, mathptmx, fixltx2e, graphicx, xspace}
\usepackage{txfonts}
  \def\apjs{ApJS} 
\def\apjl{ApJ}   \def\aap{A\&A}

\def\MsunPerYear{~M$_{\sun}$~yr$^{-1}$}

\newcommand{\Msun}{~M$_{\sun}$\xspace}
\newcommand{\percubiccm}{~cm$^{-3}$\xspace}
\newcommand{\Gasoline}{{\sc gasoline}\xspace}
\newcommand{\Ramses}{{\sc ramses}\xspace}

\begin{document}

\title{A comparison of black hole growth in galaxy mergers with \Gasoline and \Ramses}
\titlerunning{Galaxy merger code comparison}

\author{
J. M. Gabor\inst{1}  
\and Pedro R. Capelo\inst{2}
\and Marta Volonteri\inst{3}
\and Fr\'ed\'eric Bournaud\inst{1}
\and Jillian Bellovary\inst{4}
\and Fabio Governato\inst{5}
\and Thomas Quinn\inst{5}
}
\institute{CEA-Saclay, IRFU, SAp, F-91191 Gif-sur-Yvette, France
\and Center for Theoretical Astrophysics and Cosmology, Institute for Computational Science, University of Zurich, Winterthurerstrasse 190, CH-8057 Z{\"u}rich, Switzerland
\and Institut d'Astrophysique de Paris, UMR 7095 CNRS, Universit\'e Pierre et Marie Curie, 98bis Blvd Arago, 75014 Paris, France
\and Department of Astrophysics, American Museum of Natural History, Central Park West at 79th Street, New York, NY 10024, USA
\and Astronomy Department, University of Washington, Box 351580, Seattle, WA, 98195-1580
}
\authorrunning{Gabor et al.}

\abstract {

Supermassive black hole dynamics during galaxy mergers is crucial in
determining the rate of black hole mergers and cosmic black hole
growth.  As simulations achieve higher resolution, it becomes
important to assess whether the black hole dynamics is influenced by
the treatment of the interstellar medium in different simulation
codes.
We here compare simulations of black hole growth in
galaxy mergers with two codes: the Smoothed Particle Hydrodynamics
code \Gasoline, and the Adaptive Mesh Refinement code \Ramses.  We
seek to identify predictions of these models that are robust despite
differences in hydrodynamic methods and implementations of sub-grid
physics.  We find that the general behavior is consistent between
codes. Black hole accretion is minimal while the galaxies are
well-separated (and even as they ``fly-by'' within 10~kpc at first
pericenter).  At late stages, when the galaxies pass within a few kpc,
tidal torques drive nuclear gas inflow that triggers bursts of black
hole accretion accompanied by star formation.  We also note
quantitative discrepancies that are model-dependent: our \Ramses
simulations show less star formation and black hole growth, and a
smoother gas distribution with larger clumps and filaments, than our
\Gasoline simulations.  We attribute these differences primarily to
the sub-grid models for black hole fueling and feedback and gas
thermodynamics. The main conclusion is that differences exist
  quantitatively between codes, and this should be kept in mind when
  making comparisons with observations. However, reassuringly, both
  codes capture the same dynamical behaviours in terms of triggering
  of black hole accretion, star formation, and black hole dynamics.}

\maketitle
\label{firstpage}

\section{Introduction} \label{sec.intro}

Galaxy mergers are thought to be transformational events in galaxy
evolution.  Mergers transform stellar disks into spheroids
\citep[e.g.][]{toomre72, gerhard81, negroponte83, barnes96}.  Via
tidal torques, they tend to compress gas into the central regions of
galaxies, triggering powerful starbursts \citep[e.g.][]{sanders88,
  barnes91, mihos96}.  The increase in nuclear gas is further thought
to fuel growth in galaxies' central supermassive black holes,
resulting in active galactic nuclei \citep[AGNs, e.g.][]{sanders88, hernquist89, dimatteo05}.

Correlations between supermassive black hole mass and global galaxy
properties \citep[see][and references therein]{kormendy13} suggest a
possible evolutionary link between black hole and galaxy growth
\citep[e.g.][]{silk98, wyithe03}.  The energetic output from AGNs may
provide a physical driver for this link: in some cases, they emit
sufficient energy to heat up all the cold gas in a galaxy, and they
may trigger powerful outflows \citep[e.g.][]{crenshaw03, dimatteo05,
  rupke11, gabor14}.  AGNs sometimes power radio jets that can heat
intergalactic and intracluster gas \citep[e.g.][]{fabian00, voit05,
  randall11}.  AGNs have been invoked in many models of galaxy
evolution as a primary actor in regulating the star formation rates
(SFRs) and stellar masses of the most massive galaxies
\citep[e.g.][]{hopkins06_unified, croton06, somerville08}.

Numerical hydrodynamic simulations have led to many advances in our
understanding of galaxy mergers and black hole fueling, but we have
not yet developed a complete understanding of how results depend on
details of these models.  Different studies have used various
treatments of hydrodynamics and feedback processes (such as supernova
and AGN output), as well as different resolution.  Details of the
feedback treatment can have an important effect on black hole growth and
its impact on galaxies \citep[e.g.][]{debuhr11, wurster13, wurster13_compare, newton13}.

Differences in hydrodynamic method can also lead to differences in
simulated galaxy (and inter-galactic) properties in various contexts.
Recent work shows that Smoothed Particle Hydrodynamics
\citep[SPH;][]{lucy77, gingold77}, historically a commonly-used method
in extragalactic astrophysics for solving the equations of
hydrodynamics, is inaccurate in certain circumstances -- for example
in resolving shocks and Kelvin-Helmholtz and Rayleigh-Taylor
instabilities \citep{agertz07}.  Grid techniques, such as Adaptive
Mesh Refinement \citep[AMR;][]{berger89}, generally improve upon these
problems~\citep{agertz07}, but with drawbacks including advection
errors, angular momentum conservation, and numerical diffusion
(e.g. \citealp{wadsley08, hahn10}; see discussion and references in
\citealp{hopkins14_gizmo}). More recent hybrid techniques employing a
moving mesh \citep{springel10} or no mesh at all
\citep{hopkins14_gizmo} help resolve these difficulties.

In cosmological simulations, traditional SPH as implemented in the
\textsc{gadget} code \citep{springel05} tends to allow a smaller
quantity of gas to cool and fuel galaxies than mesh-based codes,
leading to smaller gas disks \citep{keres12, vogelsberger12,
  scannapieco12}.  The apparently suppressed cooling in SPH relative
to moving mesh methods also allows the formation of more prominent hot
halos in idealized merger simulations, along with clumps and filaments
in those halos \citep{hayward14}.  This is true despite the fact that,
in cosmological SPH simulations, a smaller proportion of gas accreting
onto galaxies passes through a hot phase \citep{nelson13}.  (Note that
recent improvements to SPH have helped alleviate many of these
discrepancies; e.g. \citealt{beck15} and references therein.  Such
improvements have recently been implemented in \Gasoline as described
in e.g. \citealt{keller14}.) Despite these important differences, it
appears that differences in feedback implementations cause more
prominent changes than differences in the numerical method for
hydrodynamics \citep{scannapieco12}.  Moving forward, large
code-comparison projects like AGORA \citep{kim14} should help
elucidate these issues in the context of galaxy evolution.  With the
exception of \citet{hayward14}, the impact of different codes and
numerical techniques on supermassive black hole fueling and feedback
has scarcely been studied.

In this work, we compare high-resolution simulations of black hole
growth in galaxy mergers with two codes: the \Gasoline SPH code, and
the \Ramses AMR code.  While many authors have used SPH codes to study
black holes in idealized galaxy mergers, relatively few have used AMR
codes \citep[see e.g.][]{kim11}.  We use standard physical recipes in
both codes for star formation, black hole growth, and stellar and
black hole feedback.  In \S\ref{sec.sims} we describe the simulations.
Then we highlight similarities and differences, and attempt to explain
them, in \S\ref{sec.results}.  We conclude in \S\ref{sec.conclusion}.

\section{Simulations}
\label{sec.sims}
%
%
\begin{figure*}
\begin{centering}
\includegraphics[width=170mm]{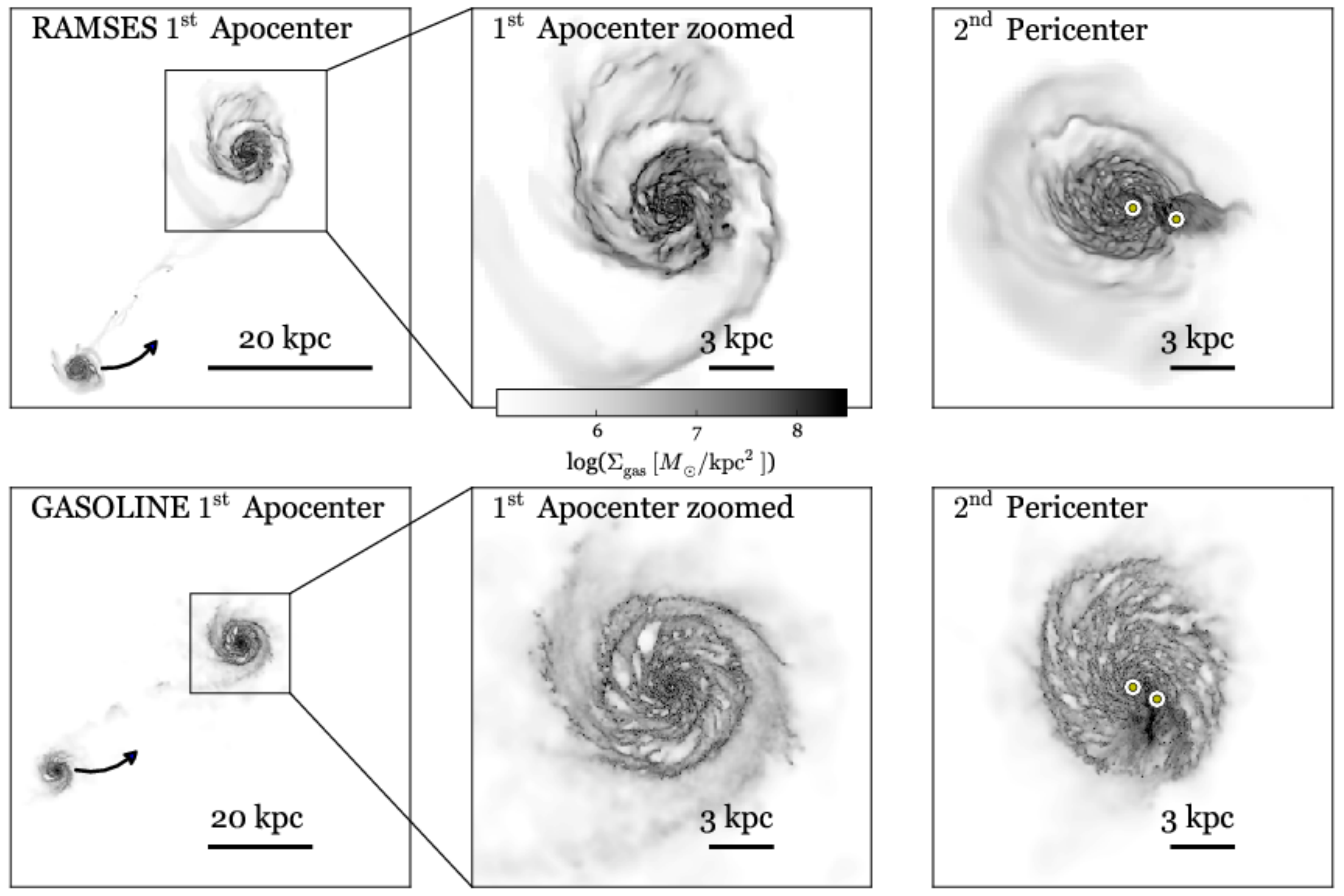}
\caption{Snapshots showing gas surface density for our 4:1 \Ramses
  ({\bf top}) and \Gasoline ({\bf bottom}) simulations.  {\bf Left}
  panels show both galaxies at first apocenter, with arrows roughly
  indicating the direction of travel of the secondary galaxy.  {\bf
    Middle} panels show a zoom-in on the primary galaxy at first
  apocenter, and {\bf right} panels show both galaxies at second
  pericenter.  In the right panels, circles mark the positions of the
  black holes (in each simulation, the more massive black hole is to
  the left).}
\label{fig.6panel}
\end{centering}
\end{figure*}
\begin{figure*}
\begin{centering}
\includegraphics[width=6.75in]{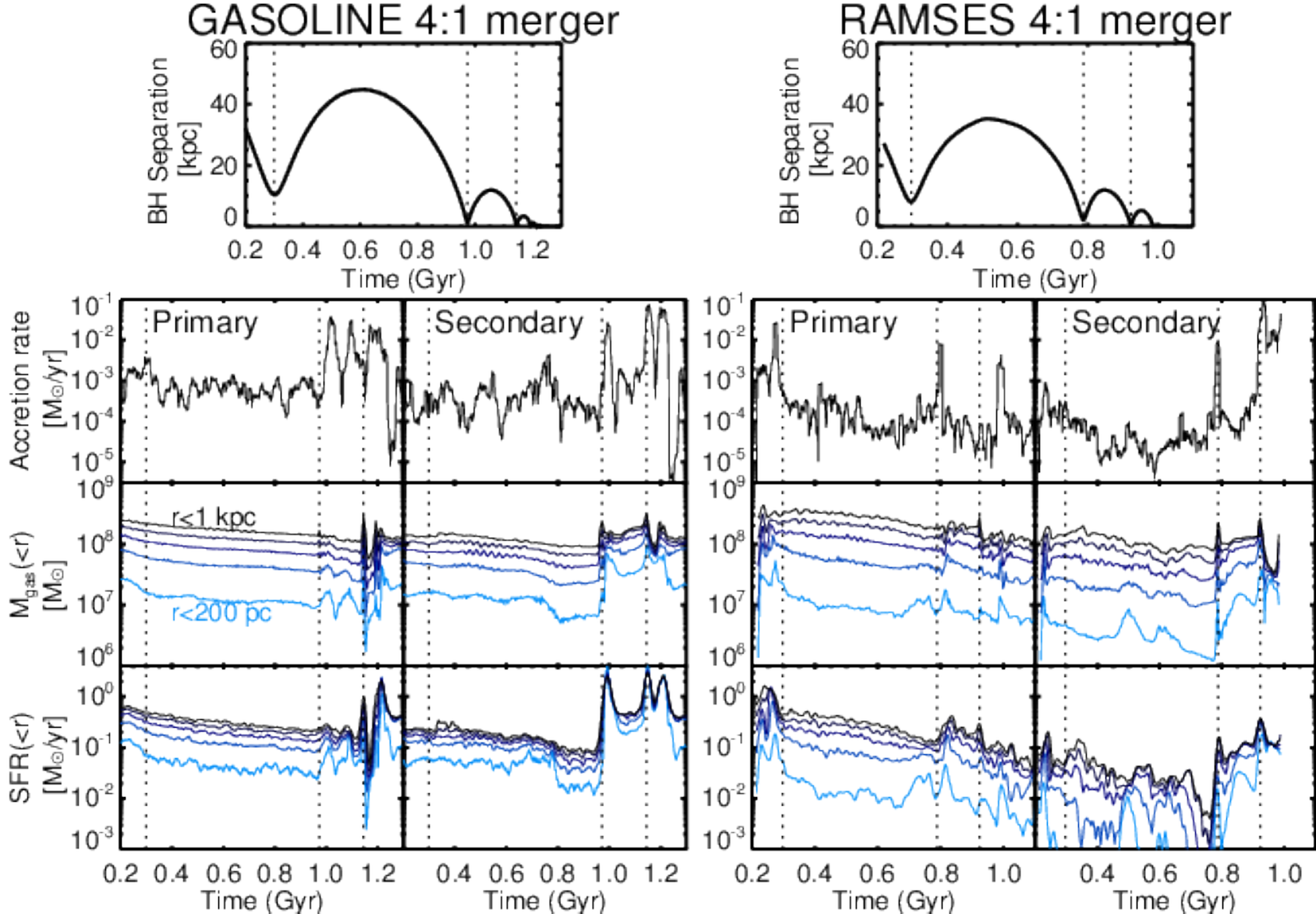}
\caption{BH separation in kpc ({\bf top}), black hole accretion rate (BHAR) in \MsunPerYear ({\bf
    second row}), gas mass in \Msun within spheres up to 1~kpc in
  radius ({\bf third row}), and SFR in \MsunPerYear within the same
  spheres ({\bf bottom row}) vs. time for a 4:1 merger with \Gasoline
  ({\bf left}) and \Ramses ({\bf right}).  For each simulation, we
  show quantities for the primary galaxy/BH on the left, and the
  secondary galaxy/BH on the right.  We show gas masses and SFRs
  within radii of 200~pc to 1~kpc, in increments of 200~pc.
  Quantities are smoothed on 10~Myr timescales.  Note that time-axes
  differ by $\sim 20$\% (see \S\ref{sec.orbits}).  Vertical dotted
  lines mark local minima in the separation between the two black
  holes in all panels.  The \Gasoline and \Ramses simulations show
  qualitatively similar results: enhanced accretion and star formation
  rates at second pericenter, and especially at 3rd
  pericenter/coalescence.}
\label{fig.compare_ts_4}
\end{centering}
\end{figure*}
For our analysis, we focus on four representative simulations of
galaxy mergers: two run with \Gasoline and two with \Ramses.  The
merging galaxies have mass ratios of 2:1 and 4:1, gas fractions of
30\%, and disks oriented coplanar with their orbits.  We have also run
some resolution tests and simulations with other mass
ratios and orbital configurations \citep[see][]{capelo15}, which give similar results.  Below
we describe numerical details of the \Gasoline simulations --
including details of the initial conditions -- then those of the
\Ramses simulations. 

\subsection{Methods: \Gasoline}
We use a subset of the suite of \Gasoline merger simulations described
fully in \citet{capelo15}.  We summarize the simulations here.
\Gasoline \citep{wadsley04}, an N-body SPH code, is based on
\textsc{pkdgrav} \citep{stadel01}, which uses a tree method to calculate gravitational
dynamics among particles.  We note that these simulations do not include recent improvements in the SPH method \citep{keller14}.

As a Lagrangian particle code, the resolution is automatically
adaptive -- the highest resolution occurs in the densest regions.
Dark matter particles, initial star particles, and initial gas
particles have masses of $1.1\times 10^5$\Msun, $3.3\times 10^3$\Msun,
and $4.6\times 10^3$\Msun (respectively), with softening lengths of
30~pc, 10~pc, and 20~pc (respectively).

\Gasoline includes models for gas cooling, star formation, supernovae and stellar winds,
and black hole accretion and feedback.  The simulations use a standard
model for gas cooling, which incorporates metal cooling but no heating by a UV
radiation background \citep{shen10}.  A temperature floor of 500~K is
imposed.

Star formation occurs as a random process \citep[see][]{katz92_sf} in gas
particles colder than 6000~K and denser than 100 particles
\percubiccm.  The SFR is calculated by assuming that a fraction
$\epsilon_*=1.5$\% of the eligible gas forms into stars per star formation
time, where the star formation time is the greater of the gas free
fall time or the gas cooling time.  With a probability consistent with
the star formation rate and timestep, a star-forming gas particle will
convert some of its mass into a collisionless star particle.

Stellar feedback includes a blast wave model for Type II supernovae
\citep{stinson06}.  In this model, supernova energy is calculated
based on the mass of young stars that should explode, and this energy
is distributed (as thermal energy) among gas particles that are
neighbors of the star particle.  To mimic an expanding blast wave, gas
cooling is turned off for a time that depends on the local gas
conditions and the feedback energy.

Nuclear black holes are treated as sink particles that accrete
surrounding gas \citep{bellovary10}.  The accretion rate onto the
black hole is calculated separately for each neighbor gas particle
based on a Bondi accretion rate \citep[cf.][]{bondi52}, and the sum of the rates from each
gas particle yields the total accretion rate.  Mass is then removed
from the gas particles in proportion to their contribution to the
accretion rate, and the mass is added to the black hole.  We note that
the Bondi accretion rate here includes a boost factor of $\alpha=3$
(see \citealt{booth09} for a discussion of boost factors).
The accretion rate is capped at $\alpha$ times the Eddington Limit.  A
fixed fraction $\epsilon_r=0.1$ of the accreted mass-energy is emitted
as radiation at each timestep, and a coupling fraction $\epsilon_f=0.001$ of the radiated energy
injected as thermal energy into the nearest gas particle.  

\subsubsection{\Gasoline simulation set up and initial conditions}
We set up mergers of two disk galaxies as in \citet{capelo15}.
Galaxies begin with parabolic orbits \citep{benson05}, initial
separations the sum of the two galaxies' virial radii, and separation
at first pericenter equal to 20\% of the virial radius of the larger
galaxy \citep{khochfar06_orbits}.  The galactic disks are coplanar with their
orbits, and both galaxies rotate in the same direction as the orbit
(prograde-prograde mergers).

Each galaxy is modeled as a dark matter halo, stellar bulge, stellar
disk, and gaseous disk \citep{springel99,
  springel05_mergers_ellipticals}, plus a supermassive black hole.
The dark matter halo follows an NFW \citep{nfw} profile up to the
virial radius, with an exponential decay beyond.  It has spin
parameter $\lambda = 0.04$ and concentration $c=3$.  The stellar bulge
makes up 0.8\% of the virial mass, and follows a \citet{hernquist90}
profile.  The galactic disk makes up 4\% of the virial mass, and
follows an exponential surface density profile.  The disk scale radius
is derived from conservation of angular momentum of the material
making up the disk.  Thirty percent of the disk mass is gas, so that
the fraction of total baryons in gas (the common observational definition of gas
fraction) is $0.3 (0.04 M_{\rm vir}) / (0.04 M_{\rm vir} + 0.008
M_{\rm vir}) = 0.25$.  Supermassive black holes with masses $2\times
10^{-3}$ times the bulge mass \citep{marconi03} are placed at the
centers of initialized galaxies.  In each simulation, the primary
galaxy has a virial mass of $2.21\times 10^{11}$\Msun, bulge mass
$1.77\times 10^9$\Msun, disk mass $8.83\times 10^9$\Msun, and disk
scale radius of 1.13~kpc.  The secondary galaxies have their masses
scaled down by factors of two and four, respectively.

Model galaxies undergo a relaxation period of 100~Myr in isolation
before they begin the merger.  During this period, the star formation
efficiency $\epsilon_*$ is gradually increased to its final value,
1.5\%, to prevent unphysical bursts of supernova feedback.  After this
relaxation, the supermassive black hole masses are reset to their
initial values, and the galaxies are placed in appropriate orbits for
the mergers.


\subsection{Methods: \Ramses}
\label{sec.method_ramses}
We have run a small new suite of merger simulations using the \Ramses
AMR code \citep{teyssier02}.  Our set up borrows many aspects from
\citet{gabor13,gabor14} and \citet{perret14}, and mostly uses standard
recipes for physical processes.  \Ramses solves the equations of hydrodynamics on the mesh,
while it treats collisionless matter (dark matter and stars) as
particles.  The effects of gravity are calculated on the mesh using a
multigrid method, and particle accelerations are interpolated using a
cloud-in-cell method.

We use the standard quasi-lagrangian mesh refinement criteria: a cell
is refined if it contains $>30$ dark matter particles, or if the mass
in the cell exceeds $5\times10^3$\Msun. In addition, we refine cells
whose mass is sufficiently large that the local gas Jeans length is
not resolved by at least 4 cell widths \citep{truelove97}.  These
refinement criteria ensure that the highest resolution is applied to
the densest regions, where star-forming clouds form.  We allow
refinement up to level 15, corresponding to a minimum cell size of
about 7.6~pc (with a box side length of 250~kpc).

The thermodynamics is treated with a standard model for gas
cooling, including metal cooling~\citep[e.g.][]{teyssier13}.  We impose an overall temperature
floor of $100$~K, as well as a density-dependent temperature floor
(the Jeans polytrope) that ensures the local Jeans length in the
smallest grid cells is always resolved by at least 4 cells
\citep{machacek01}.  The Jeans polytrope acts as an additional
pressure (or pressure floor) that prevents numerical fragmentation.
The normalization of this temperature floor depends on resolution, and
is higher for lower resolution simulations.

\Ramses includes models for star formation and supernova feedback.
Star formation occurs as a random process in cells whose gas (particle
number) density exceeds 100 H atoms cm$^{-3}$.  In cells above this
threshold density, a fraction $\epsilon_* = 1$ per cent of the gas is
assumed to form into stars per free fall time, yielding an SFR for
each cell.  A new collisionless star particle is created in the
cell with a probability based on the SFR and the
timestep.  New star particles inherit the position and velocity of the
gas cell out of which they formed, but they are decoupled and will
generally move into other cells.  

After a delay of 10~Myr, supernovae explode at the locations of newly formed
star particles.  Twenty per cent of the mass of the initial star particle
is assumed to explode as supernovae, and for each 10~\Msun of
exploding supernova, $10^{51}$~ergs of thermal energy is added to the
hosting cell.  Following \citet{stinson06} and \citet{teyssier13},
cooling is delayed for 20~Myr in the supernova-heated cell to enable
more efficient feedback.

Supermassive black holes are represented by collisionless sink
particles.  The sink particles accrete gas according to a Bondi
accretion rate \citep[cf.][]{bondi52}, where the gas density and
temperature are computed from a weighted average of all gas cells
within $4\Delta x$ (where $\Delta x$ is the smallest cell size,
\citealt{krumholz04}).  We use the standard Bondi formula, without a
``boost'' factor $\alpha$ \citep{booth09}.  The accretion rate is
capped at the Eddington limit, assuming a radiative efficiency
$\epsilon_r=0.1$.  Black hole particles merge once they pass within
$4\Delta x$ of one another.

Black hole feedback is implemented as a thermal deposition of energy
\citep{dubois10, teyssier11}.  A fraction $\epsilon_r=0.1$ of the
accreted mass is assumed to convert into radiative energy, and a
fraction $\epsilon_c=0.15$ of the radiative energy is assumed to
couple to the surrounding gas as thermal heating~\citep{booth09,
  dubois10, teyssier11}.  Following \citet{booth09}, we only inject
AGN feedback energy if it is sufficient to heat the surrounding gas to
$T_{\rm min, AGN} = 10^7$~K.  This prevents the deposited energy from
being immediately radiatively cooled from dense gas, leading to more
efficient feedback.  On timesteps where the feedback energy is
insufficient to heat the gas to $T_{\rm min, AGN}$, we store the
feedback energy to be added to that during the following timesteps.
This storage is repeated until enough energy is stored to reach
$T_{\rm min, AGN}$, at which time it is released.  We deposit the feedback
energy in gas cells within $4\Delta x$ with a weighting where colder,
denser gas acquires more of the energy \citep{gabor13}.  If the
post-injection gas temperature would exceed $T_{\rm max,
  AGN}=5\times10^9$~K, then we iteratively expand the injection radius
by 25 per cent to dilute the injection energy and lower the injection
temperature.  This maximum injection temperature prevents extremely
high temperatures which can cause computational problems
\citep{gabor13}.

As noted in \citet{gabor13}, numerous tests show that black holes
sometimes scatter from the centres of their host galaxies by hundreds
of pc.  In merger simulations, the scattering is even more pronounced.
Some of this scattering is a numerical artifact -- it essentially does
not occur in the \Gasoline simulations \citep[except on small scales,
  which can be reduced with improved dynamical friction modeling;
  see][]{tremmel15}.  The scattering in \Ramses is reduced, but not eliminated,
when a more accurate direct-summation N-body gravity solver is applied
to the sink particle \citep[cf.][]{bleuler14}.  On the other hand,
some scattering (at the $\sim 10^2$~pc level) is physically realistic
given the dynamics in the nucleus (see \S\ref{sec.stabilization} for
additional discussion).  To limit scattering, we adopt a well-known
solution of assigning a black hole dynamical mass that is much larger
than the true mass \citep[e.g.][]{debuhr10}.  For the gravity
calculation, the black hole is assigned a mass of $10^9$\Msun, while
we use the true mass for black hole-specific physics (e.g. Bondi
accretion).  This effectively keeps the black holes in their galactic
centers by increasing the dynamical friction due to stars and dark
matter.

\subsubsection{\Ramses simulation set up and initial conditions}

Initial conditions for the \Ramses simulations are generated
separately from those for the \Gasoline runs, following the
description in \citet{gabor13}. We use the same masses, density 
profiles and numbers of particles (for stars and for dark matter) as 
in the \Gasoline simulations, and initialize the positions and velocity 
distributions with the Cartesian grid code from \citet{bournaud02}. 

The initial conditions generator determines equilibrium positions and 
phase-space velocity distributions for stellar and dark matter particles, 
taking into account the gas disk contribution to the gravitational potential, 
and the gas density distribution if initialized analytically in \Ramses. 
We also must initialize gas cells outside the disk: technically, null densities
could not be handled properly. We do this by setting their density very low
 ($\sim 10^{-7}$\percubiccm) so that the initial contribution to the 
 circum-galactic halo is marginal: this halo will form from outflows 
 during the simulation. Supermassive black holes are added to the galactic 
 centers at the start of the \Ramses runs.

A necessary difference with the \Gasoline initial conditions is that the dark matter
halo must be truncated to fit within the chosen simulation box size. The truncation 
radius, depending on the halo and box sizes, is 70 to 85\% of the virial radius $R_{200}$. 
To maintain the halo truncation over time, we truncated the Maxwellian distribution of velocities 
for dark matter particles at 90\% of the local escape velocity (computed at the initial position
of each particle), otherwise the halo expands spatially over time and dynamically evaporates
from the \Ramses simulation box. When truncating the haloes, we 
made the choice to keep the same total halo mass, so at very large separations during 
the early phases of the interaction the two galaxies have the same total masses and keep
similar orbits with the two codes, and undergo their collision under nearly
 identical configurations. Yet at lower separations once the systems start 
to overlap the halo densities start to differ: haloes generated this way end up with slightly
higher masses within $R_{200}$ and smaller radii than the \Gasoline haloes (by a few 
percent typically), so short-range gravity and dynamical fraction are somewhat stronger
 in the \Ramses simulations. This will be discussed later when comparing the detailed 
 evolution of each merging system.

As in the \Gasoline case, we wish for each galaxy to undergo a
relaxation period of $\sim 100$~Myr.  We do not have a simple method
to insert a fully relaxed galaxy into a \Ramses merger simulation, so
the relaxation must occur during the main simulation.  Given our
orbital parameters, the galaxies remain quasi-isolated and undisturbed 
by tidal forces during the first $200-300$~Myr of the merger, which is sufficient to
 allow relaxation (1-2 disk dynamical times) and allow the formation of internal 
 features such as bars and spiral arms before the interaction itself occurs.

For the merger, galaxies begin with the same orbital parameters
(initial separation and velocities) and orientations
(prograde-prograde) as in the \Gasoline simulations.  As noted above,
the mergers evolve in a cubic box with a size of 250~kpc.

\section{Results}
\label{sec.results}
%
%
\begin{figure*}
\begin{centering}
\includegraphics[width=170mm]{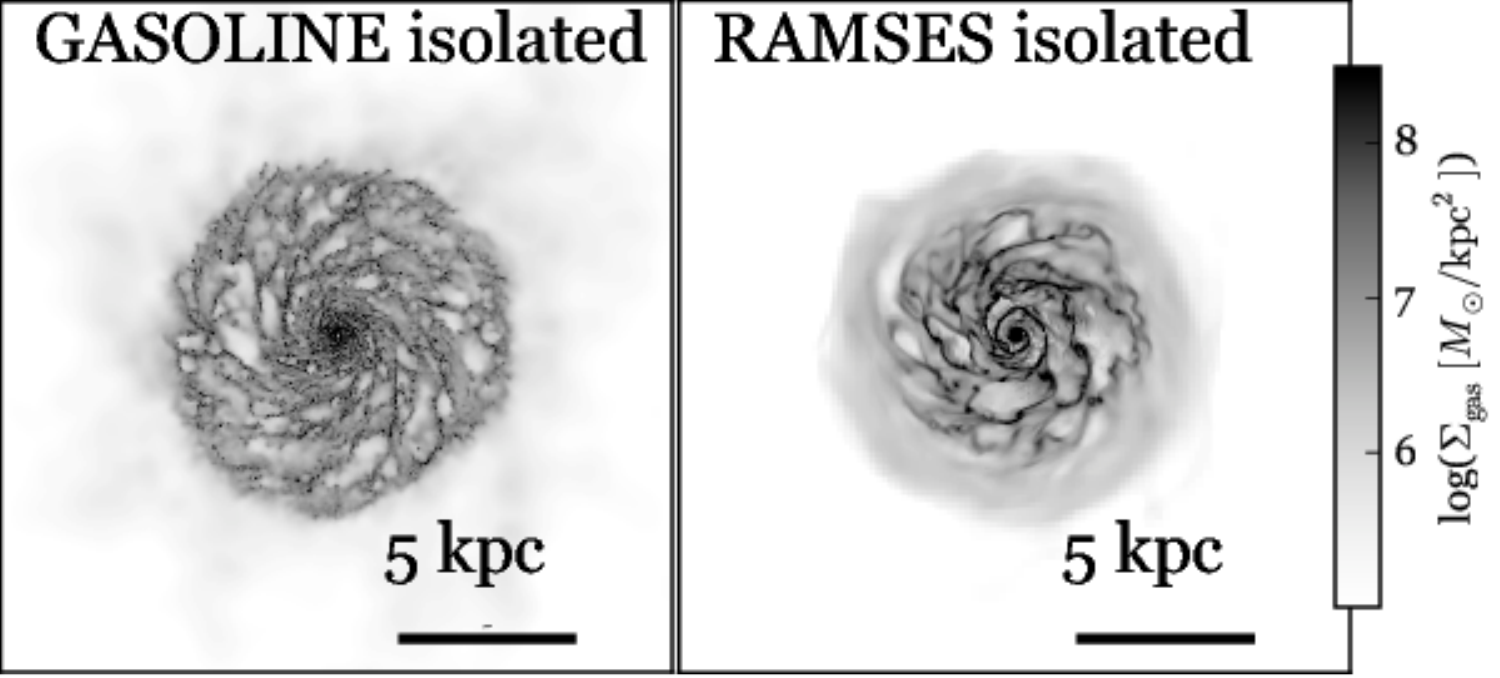}
\caption{Face-on images of gas surface density of an isolated disk
  galaxy simulation, using \Gasoline ({\bf left}) and \Ramses ({\bf
    right}).  The characteristic sizes of gas structures are larger in
  \Ramses than in \Gasoline.  This owes partly to the Jeans polytrope
  pressure floor implemented in \Ramses, and partly to differences in more
  fundamental aspects of the codes (e.g. hydrodynamic or gravity solvers).
}
\label{fig.iso}
\end{centering}
\end{figure*}
\begin{figure}
\includegraphics[width=3.25in]{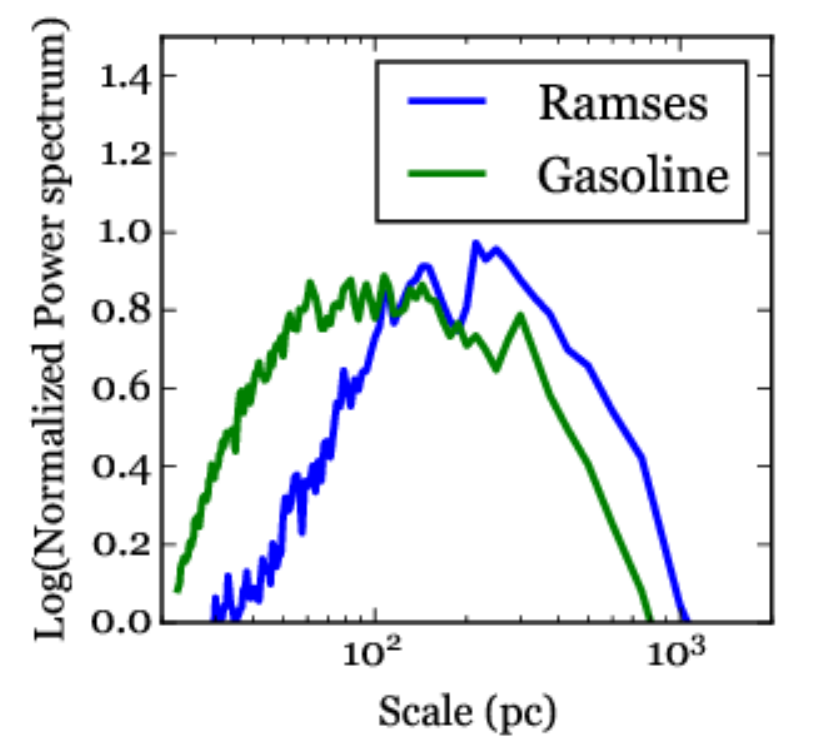}
\caption{Power spectra of gas surface density fluctuations in isolated
  \Gasoline and \Ramses galaxies.  The power spectra are normalized by
  a third-degree power law to highlight the differences between
  simulations.  \Gasoline shows more power at small scales, and less
  power at large scales, emphasizing that gas structures in our
  \Gasoline simulations tend to have smaller sizes than those in
  \Ramses.}
\label{fig.power_spec}
\end{figure}
\begin{figure*}
\begin{centering}
\includegraphics[width=130mm]{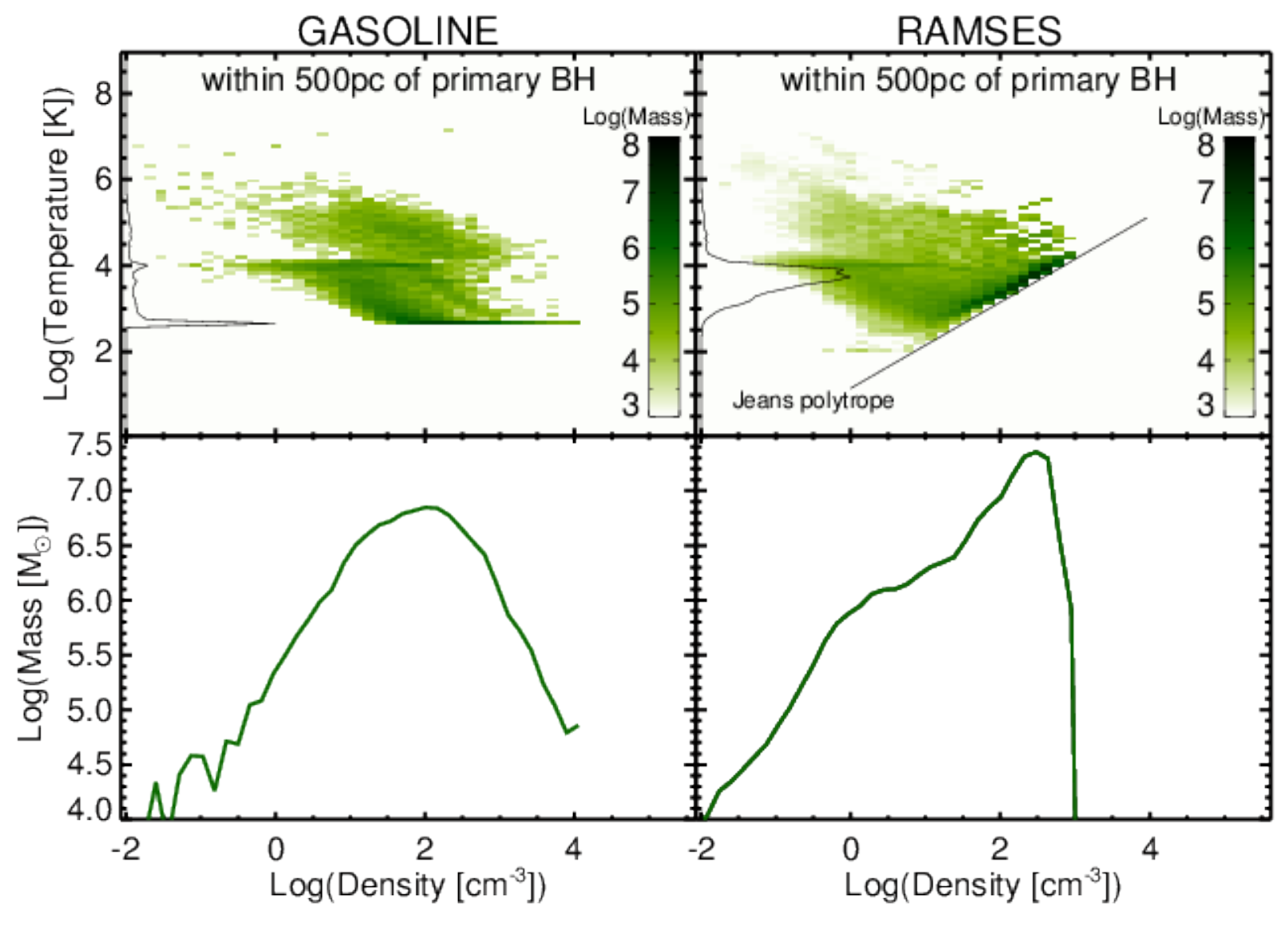}
\caption{Density-Temperature diagrams ({\bf top row})
  and density PDFs ({\bf bottom row}) for gas in \Gasoline ({\bf left
    column}) and \Ramses ({\bf right column}).  We show only gas
  within 500~pc of the primary galactic center in each simulation, and
  we use a snapshot near the maximum separation (first apocenter,
  around 500~Myr).  In the top panels, we include the temperature
  distribution along the left axis (arbitrarily normalized, and shown
  on a linear scale).  We also schematically show the Jeans polytrope
  temperature floor for the \Ramses simulation (straight black line),
  which keeps the average gas significantly hotter than in
  \Gasoline. }
\label{fig.rho_T}
\end{centering}
\end{figure*}
\begin{figure}
\includegraphics[width=3.25in]{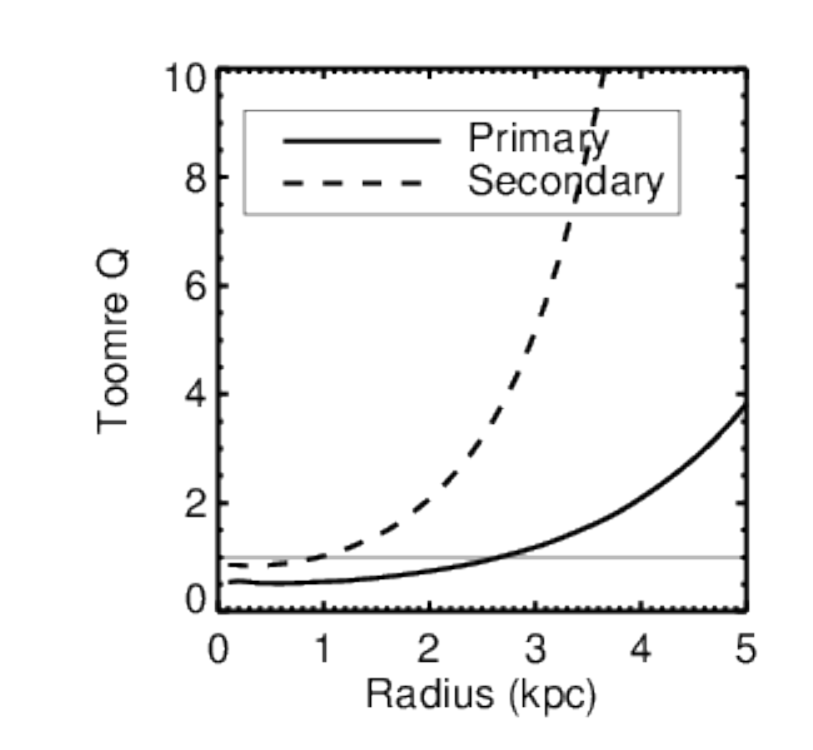}
\caption{Toomre $Q$ parameter estimated as a function of radius for
  the initial disk galaxies in a 4:1 merger in \Ramses.  A horizontal
  line marks the marginally-stable value $Q=1$.  At a gas temperature
  of $10^3$~K (which is effectively enforced by the Jeans polytrope
  temperature minimum), the secondary galaxy is substantially more
  stable than the primary, and it remains stable or near marginal
  stability at all radii.  This leads to a low SFR in the secondary.
  In the \Gasoline simulations without a Jeans polytrope, the
  secondary has higher SFRs.}
\label{fig.toomre}
\end{figure}
\begin{figure*}
\begin{centering}
\includegraphics[width=6.75in]{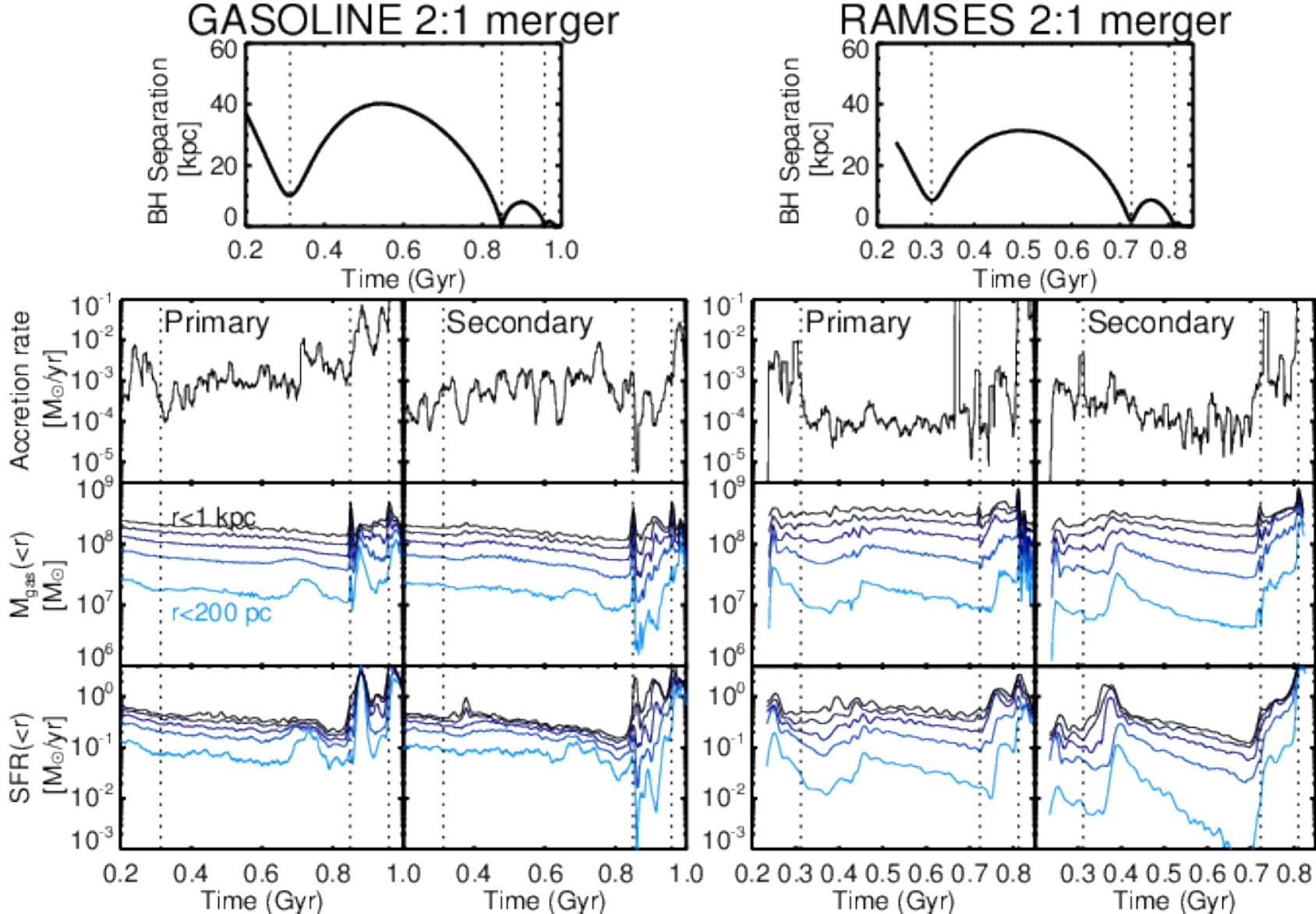}
\caption{Same as Figure \ref{fig.compare_ts_4}, but for a 2:1 merger rather than a 4:1 merger.
  BH separation in kpc ({\bf top}), BHAR in \MsunPerYear ({\bf
    second row}), gas mass in \Msun within spheres up to 1~kpc in
  radius ({\bf third row}), and SFR in \MsunPerYear within the same
  spheres ({\bf bottom row}) vs. time for a 2:1 merger with \Gasoline
  ({\bf left}) and \Ramses ({\bf right}).  For each simulation, we
  show quantities for the primary galaxy/BH on the left, and the
  secondary galaxy/BH on the right.  We show gas masses and SFRs
  within radii of 200~pc to 1~kpc, in increments of 200~pc.
 Vertical dotted
  lines mark local minima in the separation between the two black
  holes in all panels.  The \Gasoline and \Ramses simulations show
  qualitatively similar results: enhanced accretion and star formation
  rates at second pericenter, and especially at 3rd
  pericenter/coalescence.}
\label{fig.compare_ts_2}
\end{centering}
\end{figure*}

\subsection{General behaviour: common features}
Figure \ref{fig.6panel} presents face-on images of gas surface density in
our 4:1 \Ramses and \Gasoline simulations.  First we show first
apocenter, then a zoom-in of the primary galaxy at the same snapshot,
and finally the merging galaxies at second pericenter.  
In Figure \ref{fig.compare_ts_4} we compare the time evolution of
various quantities during a 4:1 merger in our \Gasoline and \Ramses
simulations.  At the top, we show the separation between the primary
and secondary black holes.  Below the separation plots, separately for
the primary and secondary, we show the black hole accretion rates (BHARs), gas
masses within 1~kpc of the black hole, and SFRs within 1~kpc of the
black hole.  For the last two, we show e.g. the gas mass within
spheres of five different radii, equally spaced from 200~pc to
1000~pc.  Careful comparison of the \Ramses and \Gasoline snapshots in Figure \ref{fig.6panel} and in the top panels
of Figure \ref{fig.compare_ts_4} indicates some differences in the galaxy orbits (e.g. in \Ramses the galaxies are
farther apart at first apocenter), and differences in gas structures
(e.g. in \Gasoline gas forms into smaller clouds).   We will address
these differences in \S\ref{sec.discrepancies}. 

Overall, however, the dynamics of the black holes is similar until we
can follow them in a consistent way. We note here that in our
\Gasoline simulations the black holes never merge, whereas in \Ramses
they merge when the galaxies coalesce, shortly after third pericenter,
once they pass within each others' accretion radii (in this case about
60~pc). Black hole mergers are implemented in \Gasoline, but we chose
to turn them off to study the dynamical
behavior~\citep{van_wassenhove14,capelo15}. This choice affects
accretion on the black holes, and therefore we refrain from commenting
on black hole growth at late times in the simulation.

The patterns in star formation and black hole accretion driven by dynamics, i.e., how 
merger-driven inflows at pericenters enhance them, are also similar, once the differences in the orbits 
are taken into account.  Indeed, one of the main conclusions of our experiment is that even when trying to make
the set-up of our simulations as close as possible in the initial conditions and in the choices of implementations, 
the way parameters are set up is intrinsically different, and therefore differences arise even in a controlled experiment.

Qualitatively, the behavior in \Gasoline and \Ramses simulations is
similar.  At first pericenter, the black holes pass within about
10~kpc of each other, but this induces little gas inflow and little
change in the BHARs or SFRs.  Between first and second pericenter, the
galaxies act essentially as though they are isolated, with relatively
steady SFRs and BHARs.  During this time the SFRs and especially the
BHARs fluctuate on short timescales -- in the BHAR case, fluctuations
of an order of magnitude are common.  These fluctuations result from
stochastic fueling of the central black hole
\citep{hopkins_hernquist06}, driven both by structure in the
interstellar medium (ISM) and by AGN feedback \citep{gabor13}.
Following \citet{capelo15}, we call this the stochastic phase of the
merger.

The action begins at second pericenter.  As the two black holes pass
within a few kpc, tidal torques trigger an increase in gas mass
within the central few hundred pc in both the primary and secondary
galaxies.  The central gas mass increase, more pronounced in the
secondary, triggers enhanced central star formation and black hole
growth.  At third pericenter $\sim150$~Myr later, the interaction
induces another burst of activity. In both simulations, this picture
(starting at second pericenter) roughly follows the well-known scenario of
merger-driven starburst and AGN fueling \citep{sanders88, barnes91, dimatteo05}.

In summary, the overall behavior of the \Ramses and \Gasoline
simulations is similar, and broadly consistent with previous studies
of galaxy mergers.  There are, however, important discrepancies
between the two simulations, which we address next. Understanding 
these differences and why they arise is important in order to extract and retain 
the results that we can consider robust.

\subsection{Quantitative discrepancies}
\label{sec.discrepancies}
\subsubsection{Orbits}
\label{sec.orbits}
The galaxy orbits differ slightly between the \Gasoline and \Ramses
runs, as implied by the top panels of Figure \ref{fig.compare_ts_4}.  In
the \Gasoline run, the galaxies move to wider separations between
first and second pericentric passages (around $t=500$~Myr), and the
second pericenter occurs at a later time ($\sim 1.0$~Gyr rather than
$\sim0.8$~Gyr in the \Ramses case).  Since the mergers are initialized
with the same positions and velocities in both simulations, we attribute these
differences to mass discrepancies.

As described in \S\ref{sec.sims}, galaxy halos in \Ramses simulations
must be truncated to fit inside a computationally reasonable box size.
When we created the initial conditions, we chose to keep the total
halo mass constant, so the truncation increases the dark matter
density by a few percent inside the truncation radius. If instead we
had kept the central density constant, then the total halo mass would
be modified and the interaction orbit would start to differ even at
large separations, before the two galaxies overlap or develop tidal
features.  Instead, our choice keeps the early-stage configuration of
the interaction relatively unchanged (see Figure
\ref{fig.6panel}). Subsequently, the higher central density of the
truncated \Ramses halo increases the gravitational forces at shorter
distances, once the two haloes start to significantly overlap.  We
attribute the faster merging in \Ramses to this difference, rather
than to effects of the codes themselves (e.g., Poisson solver type or
accuracy).  To further probe this, we estimated analytically the
timescale required to ``free-fall'' from the first apocenter to the next
encounter at $\leq$\,5\,kpc, assuming the radial profile of the haloes
and baryonic components did not evolve from the initial conditions:
the timescales are 382\,Myr for the \Gasoline initial mass
distribution, and 327\,Myr for the \Ramses one.  The timescales
measured in the simulations are about 350 and 290\,Myr, respectively:
they are both shorter than our analytic estimates (probably due to the
extra effects of dynamical friction, and/or increased mass
concentration at the first pericenter), but the relative difference of
$\simeq$15\% is fully consistent with that in the analytical estimate. Hence,
the faster merging timescale in the \Ramses merger is fully consistent
with having resulted from our choice of keeping the total halo mass constant
when a truncation is applied to the initial conditions.


\subsubsection{Black hole accretion rates}
%
%
%
During the relatively quiescent phase between first and
second pericenter, the BHARs differ significanly in the two
simulations (see Figure \ref{fig.compare_ts_4}): \Ramses BHARs fluctuate around $10^{-4.5}$\MsunPerYear,
while those in \Gasoline fluctuate around $10^{-3.5}$\MsunPerYear.  In
any case, the black hole growth during this period is small because
the accretion rates are low.  Thus the detailed level of black hole
fueling is relatively unimportant during the quiescent phase.

During the merger phase, the primary BHs show marked differences in
the BHAR.  In \Gasoline, the primary BHAR peaks sharply (by a factor
$>10$) just after second pericenter, remains elevated (with large
fluctuations) more-or-less until third pericenter, and rises again
just after third pericenter.  In \Ramses, the primary shows a brief
spike in BHAR at second pericenter, then returns to the quiescent
level even through third pericenter, until experiencing another spike
at coalescence.  The secondary BHARs, in contrast, are similar in both
simulations, showing a peak in BHAR at second pericenter and another
peak at third pericenter that lasts through coalescence.  In \Ramses,
black hole coalescence occurs $\sim50$~Myr after third pericenter,
whereas coalescence never occurs in \Gasoline.  The coalescence limits
dual black hole growth in the merger remnant.

We attribute these quantitative discrepancies between black hole
growth in \Gasoline and \Ramses mainly to differences in the black
hole fueling and feedback models.  In \Gasoline, the accretion rate
includes a boost factor of 3 to the formal Bondi rate, and the
feedback coupling efficiency is set to a relatively low value of 0.1\%
\citep[calibrated from idealized mergers and zoom
  simulations;][]{bellovary13}.  Thermal energy from feedback is dumped
into a single nearby particle at every timestep without any storage.
In \Ramses, on the other hand, the accretion rate includes no boost
factor, and the coupling efficiency is set to relatively high value of
0.15 \citep[calibrated from cosmological simulations with much lower
  resolution, and slightly different feedback prescriptions;
  see][]{booth09, dubois12_dual}.  Feedback energy is dumped into the
entire accretion region, but only if enough energy is ``stored'' to
heat the gas in that region to $10^7$~K, which increases the effective
efficiency.

Black hole growth is mostly self-regulated in these simulations,
suggesting that the difference in coupling efficiency causes more of a
difference than the boost factor.  The efficient feedback in \Ramses
keeps the gas immediately around the black hole hot and diffuse,
leading to lower accretion rates than in \Gasoline.  This effect
persists throughout both the stochastic phase and merger phase.  Owing
to the efficient feedback, bursts of accretion in \Ramses (see Figure
\ref{fig.compare_ts_4}) are short-lived and they drive only minor
BH growth.

The implementation of supernova (SN) feedback in this set of \Ramses
simulations may also contribute to suppressing black hole
growth. \citet{dubois15} find that when the SN feedback implementation
in \Ramses includes delayed cooling black hole growth is suppressed in
galaxies with bulge mass below $\sim 10^9$ \Msun, which is comparable
to the bulge mass in our runs.

This BH growth discrepancy raises questions about the choice of AGN
feedback efficiency.  Our best method to calibrate the efficiency
relies on comparing simulated black hole growth to observed BH
demographics and $M_{\rm BH}$--galaxy relations, especially $M_{\rm
  BH}-M_{\rm bulge}$.  This was done using large-scale cosmological simulations
for \Ramses, and idealized mergers and zoomed simulations for
\Gasoline~\citep{bellovary13}.  Combined with the
discrepancies in BH growth, this suggests that the optimal efficiency
could depend on resolution (and stellar feedback effects).  If so,
this calibration could be especially problematic when pushing the
boundaries of simulation resolution -- it is computationally expensive
to run several high-resolution simulations just for calibration.

In summary, different physical prescriptions for black hole fueling
and feedback dominate over differences in hydrodynamic method
\citep[cf.][]{hayward14}.  The efficient AGN feedback in \Ramses leads to lower BH
accretion rates than in \Gasoline.  The $M_{\rm gas}$ plots in Figure
\ref{fig.compare_ts_4} -- which show less gas inflow at late stages in
\Ramses than in \Gasoline -- suggest that gas dynamics play a role as
well.  In the next sections, we explore how the models for gas
thermodynamics influence the gas structure of the galaxies, which in
turn will influence the details of gas dynamics.

\subsubsection{Galaxy gas structure}
\label{sec.structure}
Figure \ref{fig.6panel} (especially the middle panels) suggests a key
difference in gas structure between our \Ramses and \Gasoline
simulations: gas forms larger, smoother structures in \Ramses.  In
order to eliminate any possible effects of the galaxy merger, we
compare these same primary galaxies but in an isolated context.  We
show images from these in Figure \ref{fig.iso}.  

In the case of \Gasoline, we show a snapshot at 200~Myr of the 4:1
merger.  This is well before the first pericenter, and thus the galaxy
appears as it would in isolation.  For the \Ramses case, we recall
that the galaxies do not undergo a separate relaxation simulation
before beginning the merger simulation, as do the \Gasoline galaxies.
Thus the galaxies relax during the first 200-300~Myr before first
pericenter.  In order to make a fair comparison with the isolated
\Gasoline galaxy, we ran a separate \Ramses simulation with an
isolated version of the primary galaxy.  We show a snapshot from this
isolated \Ramses simulation at about 250~Myrs, which is similar to the
``age'' of the \Gasoline galaxy.

Figure \ref{fig.iso} reinforces the difference in gas structure.  In
\Ramses the gas forms thick filaments and large clumps, whereas in
\Gasoline the gas forms many smaller filaments and clouds, with a more
flocculent appearance.  Moreover, dense gas structures appear to
extend to larger radii in \Gasoline, while the \Ramses outer disk is
quite smooth.

To quantify the differences in gas structure, we calculate the power
spectrum of two-dimensional spatial variations \citep[as in][]{christensen12}.  We first crop each
gas surface density map shown in Figure \ref{fig.iso} to a width of
$L\approx6$~kpc to isolate the galaxy ISM, then calculate the 2D fast
Fourier transform.  This yields a 2D image in frequency space,
$\widetilde{\Sigma}(u,v)$, where $u$ and $v$ are frequency
coordinates.  We shift the image so that zero frequency is at the
image center, and calculate the 2D power as
$P(u,v)=|\widetilde{\Sigma}|^2$.  From the 2D power image we calculate
the 1D power spectrum, $P(k)$, by calculating the average power per
pixel in a concentric set of circular annuli.  Here, $k$ refers to the
radius, in pixels, of each annulus placed on the frequency-space
image; $k$ is a spatial frequency coordinate related to spatial
wavelength $\lambda$ via $\lambda = L / k$.  Structures of size $s$
correspond to a wavelength of $\lambda = 2s$.  Finally, we normalize the power
spectrum by a power law to clarify the differences between
simulations: in Figure \ref{fig.power_spec} we show $P'(s) =
P(s) / (0.024 s^3)$.  A third-degree power law roughly fits
the shape of $P(s)$, so this normalization flattens the
trend.

Figure \ref{fig.power_spec} shows that our \Gasoline simulation has
more power than \Ramses on scales $s$ smaller than $\approx
100$~pc, and less power on scales larger than $\sim 200$~pc.  This
confirms that the sizes of gas structures in \Gasoline indeed tend to
be smaller than those in \Ramses.  Both simulations have the same
amount of power on scales $\sim 100-200$~pc, which is approximately
the width of filaments and clumps in the \Ramses simulation (i.e. the
Jeans length).  The \Gasoline simulation includes structures of this
size, but their internal fragmentation leads to the enhanced power on
smaller scales.

In the following section, we argue that the relative excess of gas
structure on large scales in \Ramses is due to the implementation of a
Jeans polytrope, the pressure floor which stabilizes the gas (which is
absent from the \Gasoline simulations).  


\subsubsection{Stabilization by the Jeans polytrope}
\label{sec.stabilization}
In Figure \ref{fig.rho_T}, we show density-temperature
diagrams and density distributions in both \Gasoline and \Ramses.  We
show only gas within 500~pc of the primary BH at a timestep during the
stochastic phase, between first and second pericenter.  This gas is
representative of gas in the disk of the galaxy.

We also schematically show the Jeans polytrope in the \Ramses density-temperature
diagram.  By construction, gas in \Ramses is not allowed to fall below
this line.  In contrast, a significant amount of gas in \Gasoline
exists at lower temperatures and higher densities than imposed by this
floor in \Ramses.  At the \Gasoline temperature floor of 500~K, gas at
the threshold for star formation ($100$\percubiccm) is resolved by
$\approx 64$ gas particles (with poorer relative resolution with
increasing gas density).  We note that \Gasoline includes an option to
use a Jeans polytrope, but it frequently goes unused in galaxy
formation simulations (as here) -- mainly because the cold dense gas to which it
applies is always star forming gas, which is treated with a sub-grid star
formation model.

The thermal pressure imposed by the effective density-dependent temperature floor in
\Ramses helps stabilize the galaxy's gas disk against gravitational
collapse.  Gas does not reach densities higher than about
$10^3$\percubiccm, whereas the high-density tail in \Gasoline, where
the temperature floor is independent of density, reaches
$10^4$\percubiccm (bottom panels of Figure \ref{fig.rho_T}).  The gas
temperature distribution (along the left axis of the density-temperature diagram)
in \Ramses has a broad peak around $10^4$~K, while in \Gasoline it
shows a bimodal structure: a small peak around $10^4$~K, and a larger
peak around $500$~K (the temperature floor).  In both simulations a
substantial portion of the gas reaches the temperature floor, but in
\Ramses the temperature floor is set by the Jeans polytrope.  

The higher typical gas temperature in \Ramses implies a more
pressurized ISM, and a larger Jeans length (and Jeans mass).  The
larger Jeans length naturally leads to larger collapsed structures,
giving rise to the larger filaments and gas clumps seen in Figures
\ref{fig.6panel} and \ref{fig.iso}.  This is reflected by the relative
enhancement of power at scales $\gtrsim 200$~pc in the gas power
spectrum (Figure \ref{fig.power_spec}).  The higher temperature in
\Ramses also affects the overall disk stability, which we address in
\S\ref{sec.SFRs}.

To test whether the Jeans polytrope is the main driver of differences
in gas structure, we ran \Ramses simulations of an isolated galaxy
without the Jeans polytrope, and repeated the power spectrum analysis
of \S\ref{sec.structure}.  In this case, the power discrepancy on
scales $\gtrsim 200$~pc between \Ramses and \Gasoline disappears (not
shown).  This implies that the Jeans polytrope is enhancing
large-scale power by increasing the Jeans length.  The relative
deficit of small scale power ($\lesssim 200$~pc) in \Ramses, on the
other hand, persists even without the Jeans polytrope pressure floor
(and also with changes to feedback models).  The small-scale
discrepency thus seems to arise from more fundamental aspects of the
simulations: numerical diffusion between gas cells in \Ramses could
smooth out small-scale perturbations, or differences in the gravity
solvers could lead to different levels of clumping.

Large gaseous clumps in \Ramses
can have masses $>10$ times the black hole mass.  The largest of these
clumps may also contribute to the black hole scattering discussed
in \S\ref{sec.method_ramses}.  We have performed a simple analytical estimate of the
scattering induced by the clumps on the black hole motion on one
\Ramses output.  We calculated the change in the black hole's velocity
both due to distant encounters \citep{binney87} and assuming a single
impulsive scatter that conserves energy and momentum, thus obtaining a
lower and upper limit to the dynamical effect.  In the case we
analyzed the ensemble of the clumps was not producing any strong
effect, but a single large clump with a mass 200 times larger than the
black hole, and very close to it ($\sim 60$~pc), could, in principle,
have imparted a kick sufficient to displace the BH from the center.

\subsubsection{SFRs}
\label{sec.SFRs}
Overall SFRs during the stochastic phase are similar in both
simulations.  In both cases, the total SFRs are dominated by those in
the primary galaxies.  This dominance is more pronounced in the
\Ramses simulation, as the SFR of the secondary galaxy is quite low
(see Figure \ref{fig.compare_ts_4}).  Notably, the gas mass within
1~kpc is similar for the two simulations, implying that \Ramses has a
lower star formation efficiency (SFR$/M_{\rm gas}$).

We attribute the abnormally low SFR in the secondary \Ramses galaxy --
but not in the primary galaxy -- to the pressure support provided by
the Jeans polytrope.  As described above, the Jeans
polytrope acts as a minimum temperature that ensures the Jeans length
is always resolved.

From Figure \ref{fig.rho_T}, the \Ramses gas density distribution
peaks around $\sim 10^2$~cm$^{-3}$, at which density the temperature floor is
around $10^3$~K.  We can quantify the stability imposed using an
analysis of the Toomre $Q$ parameter \citep{toomre64}.  In a thin
galactic disk, the value of $Q$ quantifies the stability of the disk
against gravitational collapse: small values indicate gravitational
collapse, large values indicate stability, and $Q\approx 1$ indicates
marginal instability.  We estimate $Q$ as a function of radius using:
\begin{equation}
Q = \frac{\sqrt{2} v_{\rm circ} c_s}{\pi G r \Sigma}
\end{equation}
where $v_{\rm circ}/r$ is the orbital frequency, $v_{\rm circ}(r)$ is
the circular velocity at radius $r$, $c_s$ is the gas thermal sound
speed, $G$ is the gravitational constant, and $\Sigma$ is the gas
surface density in the disk at $r$.\footnote{This
  estimate is accurate for flat rotation curves, for which the
  epicyclic frequency is $\kappa = \sqrt{2}\Omega = \sqrt{2} v_{\rm
    circ}/r$. It is an under-estimate of $Q$ in the central kpc or so
  of our simulations, where the rotation is rising.}  We calculate
$c_s$ assuming a gas temperature of $10^3$~K, which is roughly the
limit imposed by the Jeans polytrope in \Ramses.

Figure \ref{fig.toomre} shows $Q$ as a function of radius for the
initial galaxies in the 4:1 merger. The primary galaxy is
Toomre-unstable within a radius of $\sim2.5$~kpc.  The smaller,
secondary galaxy is significantly more stable than the primary, and
$Q$ barely reaches below 1 (and only in the central kpc).  Thus, the
Jeans polytrope stabilizes the smaller gas disk, inhibiting
gravitational collapse and star formation.

This stabilization may also help explain why the SFRs near the
secondary black hole during the merger phase (after second pericenter)
are lower in \Ramses than those in \Gasoline.  The amount of gas
within 1~kpc of the secondary BH is similar in both simulations, but
the SFR is lower in \Ramses (see Figure \ref{fig.compare_ts_4}).
Another contributor to the stability at smaller radii could be the
presence of the artificially-massive black hole, which would boost the
circular velocity and therefore $Q$.  In the primary galaxy, however, we see little
evidence for a suppression of the SF efficiency in the central regions
of the \Ramses simulation relative to the \Gasoline simulation,
suggesting that the extra-massive black hole plays little role.
Finally, efficient AGN feedback may help maintain the gas at higher
temperatures, suppressing the SF efficiency.

Near the primary BH, at and after third pericenter, the gas inflow to
the central kpc is more significant in \Gasoline than in \Ramses.
This leads to a larger gas mass and stronger burst of SF in \Gasoline
than in \Ramses (see Figure \ref{fig.compare_ts_4}).  The origin of
this difference is unclear.  It could result from differences in the
dynamics around the black holes, or from the efficient AGN feedback in
\Ramses that evacuates gas from the nuclear regions.

Another possible driver of differences in SFR could be the supernova
feedback recipe.  In \Gasoline, gas heated by supernovae is not
allowed to cool for a time that depends on local gas conditions. In
\Ramses, the cooling delay is fixed at 20~Myr.  Supernova feedback
helps regulate star formation, so different SFRs could arise from
these different implementations.  The primary galaxies have similar
SFRs during the stochastic phase of our simulations, so the different
supernova feedback seems to have a minor effect.

In summary, the two simulations have similar SFRs overall, but the the
\Gasoline simulation shows more SF in the secondary galaxy and during
the merger phase than does the \Ramses simulation.  We attribute these
differences to stabilization due to the Jeans polytrope in \Ramses.
This temperature floor increases the gas stability against collapse,
suppressing star formation.

\subsection{A 2:1 merger}
We show the results of a 2:1 galaxy merger in Figure
\ref{fig.compare_ts_2}.  Broadly, the behavior is similar to that in
the 4:1 galaxy merger.  The black holes undergo a relatively quiescent
stochastic phase of growth lasting until second pericenter, at which
point gas inflows trigger more activity.  The star formation rates
follow a similar broad pattern: steady star formation until second
pericenter, then a boost triggered by gas inflows.

In the \Ramses simulation, the BH accretion rates during the
stochastic phase are again lower than in \Gasoline, owing to
differences in the supernova and AGN feedback model.  The \Ramses primary BH shows
little enhancement of accretion at second pericenter, while the
\Gasoline primary shows a burst just after second pericenter.  At
third pericenter, both simulations show a burst of accretion.  The
secondary BHs show the opposite trend: the \Gasoline secondary shows
little enhancement at second pericenter, and the \Ramses secondary
shows a brief burst at second pericenter.

The star formation rates are broadly consistent between the
simulations.  The secondary galaxy in this 2:1 merger is sufficiently
unstable (owing to higher gas surface densities) in \Ramses to have
comparable SFR during the stochastic phase to that in \Gasoline.  Both
simulations show SFR enhancements between first and second pericenter,
but these occur at different times and locations in the two
simulations.  The \Gasoline primary shows an SFR enhancement about
200~Myr before second pericenter; the \Gasoline secondary shows an
unusual SFR peak at around 1~kpc within 100 Myrs after first
pericenter, then a weak central burst about 200~Myrs before second
pericenter.  In \Ramses, both the primary and secondary galaxies show
SFR bursts about 100~Myr after first pericenter, but the secondary one
is much stronger.  SFR enhancements during the merger phase again
appear stronger in the \Gasoline simulation, but the differences are
less pronounced than in the 4:1 simulation.

\section{Conclusion} 
\label{sec.conclusion}

We have compared galaxy merger simulations including black hole growth
with \Ramses, an AMR code, and \Gasoline, an SPH code, to find what is robust
against differences in the codes and their physics implementations. We ran a small
suite of new \Ramses simulations and used \Gasoline simulations from a
suite described in \citet{capelo15}.  In both cases, the simulations
had a spatial resolution of $\sim 20$~pc and mass resolution of $\sim
5\times 10^3$\Msun, and included standard sub-grid recipes for gas
cooling, star formation, supernova feedback, and black hole fueling
and feedback. 

The simulations show general agreement in their dynamical behavior,
and the results are broadly in line with the long-standing picture of
galaxy merger simulation evolution \citep[cf.][]{dimatteo05,
  hopkins06_unified}.  Black hole growth is relatively slow while the
galaxies are well-separated, including the time between first and
second pericenter.  During this phase, black hole growth is driven by
stochastic accretion \citep{hopkins_hernquist06, gabor13, capelo15}, just as
for isolated galaxies.  Once the galaxies' effective radii overlap at
second pericenter, tidal torques trigger gas inflows toward the
nucleii.  The activity is strongest when the galaxies finally coalesce
(beginning at third pericenter).  The inflowing gas, driven to higher
densities, initiates peaks in the SFR and the black hole accretion
rate.

While the general behavior of the galaxy mergers is similar, there are
quantitative differences between simulations run with different codes.
We attribute most of these discrepancies to differences in sub-grid
cooling and feedback models, but effects of the hydrodynamic methods
(SPH and AMR) or even gravity solvers (gravity tree and Particle Mesh)
may play a role.  In \Ramses simulations, nuclear black holes accrete
significantly less than in \Gasoline, both while the galaxies are
well-separated and during coalescence.  This arises due to a higher
feedback efficiency in \Ramses, which regulates the black hole growth
at a lower level.  Furthermore, the more efficient black hole feedback
ensures that central star formation is lower in \Ramses during the
merger and coalescence.  Another difference between codes is the ISM
gas structure: in \Gasoline gas forms small filaments and collapses
into tiny dense knots, whereas in \Ramses gas forms thick filaments
and large clumps.  Gas structure differences owe partly to the
inclusion of a Jeans polytrope pressure floor in \Ramses, though at
the smallest scales they may be related to numerical methods.

While differences in hydrodynamic method are important in some regimes
of galaxy evolution models \citep[e.g.][]{hayward14}, we conclude that
the most important differences arise in the treatment of baryonic
physics \citep[cf.][]{scannapieco12}.  Future work will better address
the effects of hydrodynamic method by making the baryonic physics
treatments as similar as possible, but due to differences in code
structure, this is not always possible. 

Our study underscores that simulation codes that successfully
reproduce a range of observables are robust at predicting the same
general physical behaviour in galaxy mergers, however, one should be
careful when comparing quantitatively one particular simulation to one
given observable, as these can be model-dependent. The trends, on
the other hand, are robust, and can be used to obtain physical
insight.

 
 

\section*{Acknowledgements} 
FG acknowledges support from grant AST-1410012 and NASA ATP13-0020.  TQ acknowledges support from NSF award AST-1311956.

\bibliographystyle{aa} 

\bibliography{paper}


\label{lastpage}

\end{document}